\documentclass[12pt]{article}
\usepackage{multirow}
\usepackage{epsfig, setspace, pmgraph}
\usepackage{subfigure}
\usepackage{graphicx}
\usepackage{afterpage}
\usepackage{amsmath, amsfonts, amssymb, upgreek}
\usepackage{dsfont,courier}
\usepackage{verbatim}
\usepackage{longtable}
\usepackage{color,soul}
\begin{document}
\pagestyle{empty}
\begin{center}
    \Large{Multiaxial Fatigue Behaviour of A356-T6}\\
    \vspace{.5 in}
    \normalsize{M. Roy$^{\text{a,*}}$, Y. Nadot$^{\text{b}}$, D. M. Maijer$^{\text{a}}$, G. Benoit$^{\text{b}}$}\\
    \vspace{.25 in}
    $^{\text{a}}$\emph{\small{Dept. of Materials Engineering, The University of British Columbia, Vancouver, BC, Canada V6T 1Z4}}\\
    \vspace{0.125 in}
    $^{\text{b}}$\emph{\small{Institut PPRIME - CNRS - Universit\'{e} de Poitiers - ENSMA - UPR 3346 - D\'{e}partment M\'{e}canique des Mat\'{e}riaux - T\'{e}l\'{e}port 2 - 1 Avenue Cl\'{e}ment Ader - BP 4019 - 86961 FUTUROSCOPE CHASSENEUIL Cedex - France}}\\
\end{center}

\hrule
\begin{abstract}
\noindent Aluminum alloy A356-T6 was subjected to fully reversed cyclic loading under tension, torsion and combined loading. Results indicate that endurance limits are governed by maximum principal stress. Fractography demonstrates long shear mode III propagation with multiple initiation sites under torsion. Under other loadings, fracture surfaces show unique initiation sites coincidental to defects and mode I crack propagation. Using the replica technique, it has been shown that the initiation life is negligible for fatigue lives close to $10^6$ cycles for combined loading. The natural crack growth rate has also been shown to be comparable to long cracks in similar materials.
\end{abstract}
\hrule
\vspace{0.1 in}
\footnotesize
\textsuperscript{*}Corresponding author, Tel. +1 604 827 5346; Fax +1 604 822 3619\\
\emph{Email addresses:} majroy@interchange.ubc.ca (M. J. Roy), yves.nadot@lmpm.ensma.fr (Y. Nadot), daan.maijer@ubc.ca (D. M. Maijer), guillaume.benoit@lmpm.ensma.fr (G. Benoit)
\normalsize
\clearpage
\section*{Nomenclature}
\begin{description}
\item[$a$] Crack length
\item[$A$] Average fatigue limit in torsion
\item[$\sqrt{\text{area}}$] Defect size parameter defined as the square root of a defect cross-sectional area
\item[$C$] Average tensile fatigue limit
\item[$J_{1_\text{max}}$] Maximum hydrostatic stress
\item[$J_{2,a}$] Amplitude of the second invariant of the deviatoric stress
\item[$K$] Stress intensity factor
\item[$\Delta K$] Positive load cycle stress intensity range
\item[$\Delta K_{\text{eff}}$] Effective stress range for crack growth
\item[$K_{\text{op}}$] Stress intensity required to open a crack
\item[$N_f$, $N$] Number of cycles to failure, number of cycles
\item[$R$] Load ratio where $R=\sigma_{\text{min}}/\sigma_{\text{max}}$
\item[$\alpha$] Crack size coefficient
\item[$\beta$] Crack size coefficient
\item[$\lambda_2$] Secondary dendrite arm spacing
\item[$\rho$] Crossland parameter
\item[$\sigma_{1_\text{max}}$] Maximum principal stress
\item[$\sigma_{a}$] Tensile load amplitude
\item[$\sigma_{f}$] Specimen-specific tensile fatigue limit
\item[$\tau_{a}$] Torsion load amplitude
\item[$\tau_{f}$] Specimen-specific torsional fatigue limit
\item[$\phi$] Stress intensity factor coefficient for surface cracks
\end{description}

\doublespacing
\section{Introduction}\label{sec:intro}
Cast aluminum alloys have been adopted over cast iron in many different applications to realize the benefits of weight reduction and part consolidation. Their relative low fatigue resistance can be an obstacle for structural components.  A direct link between microstructure and fatigue resistance has been demonstrated by many authors \cite{Gao.04,Skallerud.93,Zhang.00,McDowell.03,Zhu.07,Wang.01}. Coarse microstructures, as characterized by large secondary dendrite arm spacing ($\lambda_2$), lead to diminished fatigue resistance.  In defect-free material, the first cracks are known to initiate either inside the primary $\alpha$-Al phase \cite{Brochu.10}, at Fe-rich intermetallic particles \cite{Gao.04}, or in the secondary phase\cite{McDowell.03}. In material containing defects, fatigue behaviour is affected by gas pores, shrinkage porosity, and/or oxides.  The studies on the high-cycle fatigue behaviour of cast aluminum primarily report that defects are of primary importance for a fatigue life assessment. While it is clear that defects are observed at the initiation point on fracture surfaces, very few studies have identified the critical defect size that diminishes the fatigue limit. Brochu \cite{Brochu.10} suggested that the critical defect size is 150 $\upmu$m for A357 and McDowell \cite{McDowell.03} found a critical defect size of 200 $\upmu$m for A356-T6. With few exceptions, all of the preceding studies considered the fatigue behaviour under tensile loading. The multiaxial fatigue behaviour of A356-T6 was studied by De-Feng et al. \cite{De-Feng.08} with thin-walled tubular specimens but under loading conditions leading to very low cycle fatigue; as such these results are not directly applicable to high cycle fatigue (HCF). Fan et al. \cite{McDowell.03} analyzed fatigue data under tension, torsion and combined loading and found that the equivalent von Mises deformation was able to correlate the results. However, these tests were conducted using deformation control leading to the possibility of a different material response.

The aim of the current study is address the scarcity of multiaxial HCF data for A356-T6. Tension, tension-torsion and torsion fatigue tests were performed for fully reversed loading and fatigue mechanisms were analyzed through fracture surface observation and replica studies. Basic fatigue criteria are also compared to discuss the multiaxial behaviour of the material.
\section{Material and Experimental Methodology}\label{sec:Material}
The material employed in this study was strontium modified A356 (Al-7Si-0.3Mg) in the T6 condition that was taken from the melt supply of a North American aluminum alloy wheel manufacturer. The typical chemical composition of this material is given in Table \ref{table:composition}.  While all specimens came from castings made with permanent steel dies, the majority of specimens came from a wedge-shaped casting produced for this investigation, and a lesser number were cut directly from an automotive wheel removed from the production stream. The wheel casting was actively cooled during solidification while the wedge casting was passively left to cool. Thus, these two castings represent a wide range of solidification conditions that lead to a range of defects and microstructure in the specimens cut from them. Considering the range of conditions, the fatigue behaviour characterized in this work is directly applicable to commercial castings.

\begin{table}[h!]
\centering
\caption{A356 composition in wt-\%\label{table:composition}}
    \begin{singlespacing}
    \begin{tabular}{lllll}
    \hline
    \textbf{Element} & Si & Mg & Na & Sr\\
    \textbf{Range (wt-\%)} & 6.5-7.5 & 0.25-0.4 & $\sim$0.002 & $\sim$0.005\\
    \hline
\end{tabular}
\end{singlespacing}
\end{table}

\subsection{Material preparation}
The motivation for using a wedge-shaped casting (Fig 1a) was to create a cooling rate gradient  during solidification that varied with height in the casting. A more refined microstructure was expected at the base of the casting where the cooling rate was the highest and a coarser microstructure occurred at the top where the cooling rate was the lowest. One half of the wedge was sectioned to produce tension-torsion fatigue specimens, while the other half was used to produce a mixture of solid tension, torsion and tension-torsion fatigue specimens (Fig 1b). Solid specimens were employed as opposed to tubular specimens in order to retain sampling within specific microstructural regions of the wedge. Metallographic specimens were taken such that there was at least two samples per height where fatigue specimens were cut in the wedge (Fig 1a). Additional tension-torsion fatigue and metallographic specimens were sectioned from the spokes of a Low Pressure Die Cast (LPDC) automotive wheel. All specimens were subjected to a T6 heat treatment: solutionized at 538$^{\circ}$C for 3 hours, quenched in 60$^{\circ}$C water, and artificially aged at 150$^{\circ}$C for 3 hours.

\begin{figure}
\centering
\includegraphics[width=0.5\linewidth]{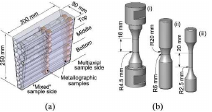}
\caption{Wedge casting dimensions and specimen locations (a) and (b) the different types of fatigue specimens employed: tension-torsion (i), torsion (ii) and  tension (iii) type.}
\end{figure}

\subsection{Microstructure}\label{sec:microstructure}
Aluminum alloy A356 has an as-cast microstructure consisting of a primary dendritic structure, $\alpha$-Al, filled with Al-Si eutectic phases. As the cooling rates decrease and solidification time increases, the dendritic structure coarsens resulting in larger secondary dendrite arm spacings. The T6 heat treatment serves to modify and refine the Al-Si eutectic structure. Depending on the casting conditions, the as-cast material may contain a variety of defects include gas and shrinkage porosity, and/or oxides which remain after heat treatment. Fig 2 shows optical micrographs of the typical microstructure. Metallographic analysis did not present any evidence of Fe-rich intermetallic phases known to affect fatigue properties.

\begin{figure}
\centering
\includegraphics[width=0.5\linewidth]{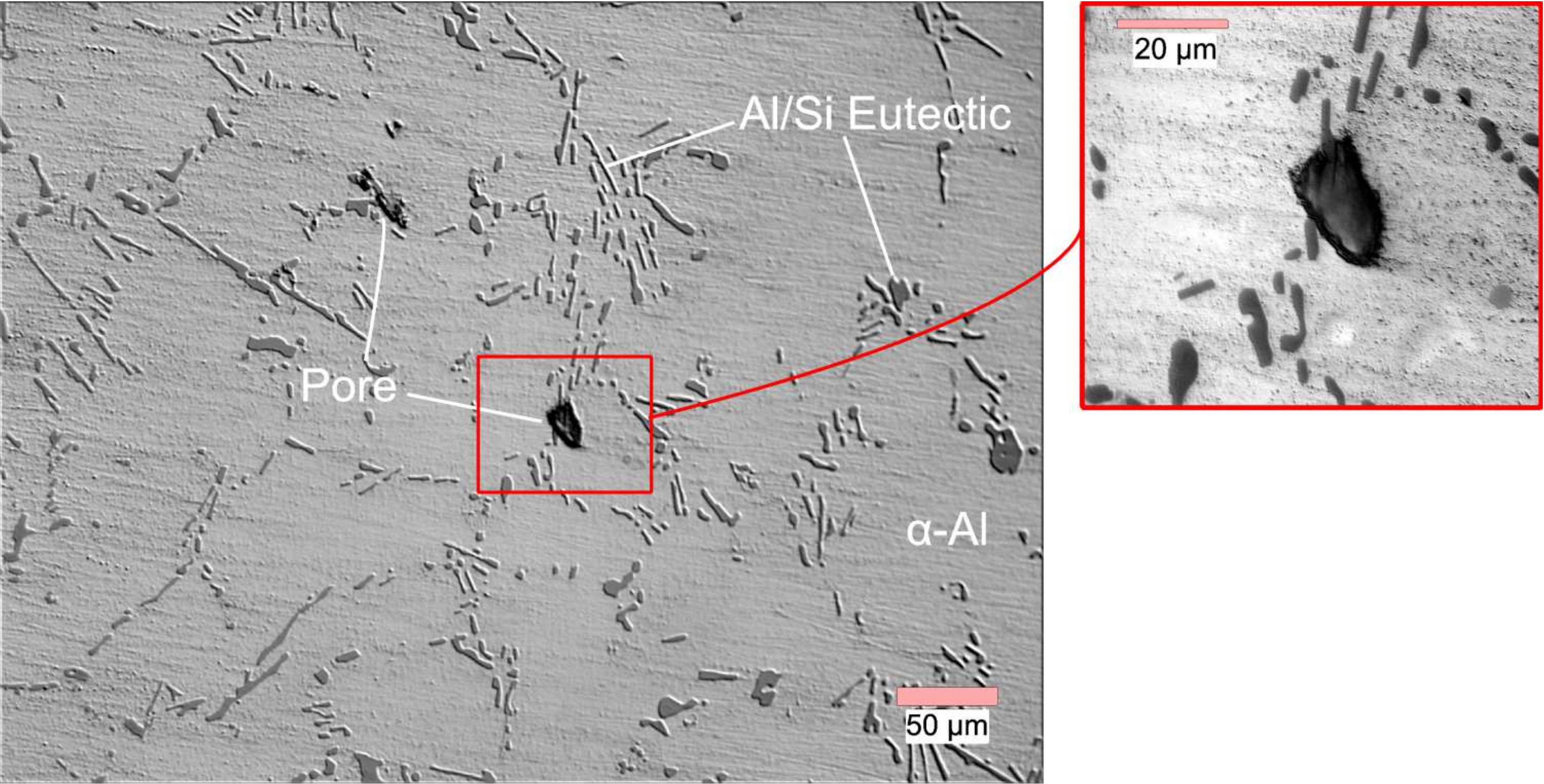}
\caption{Typical A356-T6 microstructure showing $\alpha$-Al, Al-Si eutectic phases and porosity.}
\end{figure}

The secondary dendrite arm spacing ($\lambda_2$) and porosity were measured on each of the metallographic specimens from the wedge and wheel with an optical microscope and the Clemex Vision PE software. Each measurement was made on a series of images representing a composite area of over 250 mm$^2$. The measured microstructure data is presented in Table \ref{table:SDAS_pores} with data arranged according to height from the bottom of the wedge. The specimens were also grouped as four families of specimens: one for the wheel, and one each for the bottom, middle and top regions within the wedge.  Due to the structure being equiaxed, $\lambda_2$ was taken to be the average distance between the centers of secondary dendrite arms based on a minimum of 60 measurements per sample. Equivalent pore diameter was calculated by equating the area of measured pores to equivalent circular geometry.

\begin{table}[h!]
\centering
\caption{Secondary dendrite arm spacing and porosity measurements for material tested.\label{table:SDAS_pores}}
    \begin{singlespacing}
    \begin{tabular}{lllll}
    \hline
    \multirow{2}{*}{Family} & \multirow{2}{*}{Height (mm)} & \multirow{2}{*}{$\lambda_2$ ($\mu$m)} & \multicolumn{2}{c}{Mean Porosity} \\
                            &                                       &   & Area (\%) & Max. Dia. ($\upmu$m)\\
    \hline
Wheel (W)	& N/A&36.7 $\pm 8$& 0.0603&58.6\\
\hline
\multirow{3}{*}{Wedge Bottom (B)}	& 28.75&39.5  $\pm$7.6& \multirow{3}{*}{0.1237}&36\\
&  58.75&39.7    $\pm$9.1 & &61\\
&  88.75&47.6   $\pm$14.0 & &124\\
\hline
\multirow{3}{*}{Wedge Middle (M)}	& 118.75&   57.2   $\pm$17.9 & \multirow{3}{*}{0.1244}&105\\
&144.73&58.5    $\pm$21.0 & &45\\
&174.73&59.7   $\pm$21.2& &93\\
\hline
\multirow{2}{*}{Wedge Top (T)}	&204.73&62.6   $\pm$21.6& \multirow{2}{*}{0.1232}&48\\
&234.73&72.2   $\pm$28.2& &126\\
\hline
\end{tabular}
\end{singlespacing}
\end{table}

The measured $\lambda_2$ increases from the bottom (39.5 $\upmu$m) to the top  (72.2 $\upmu$m) of the wedge, coinciding with the variation in cooling rate imposed by the casting practice.  The $\lambda_2$ measured in the wheel specimen (36.7 $\upmu$m) was similar to that measured at the bottom location in the wedge. The area percent porosity throughout the wedge was uniform (~0.12 \%) and approximate double that measured in the wheel (0.06 \%). The reduced porosity content is consistent with the in-line degassing practices employed by the wheel manufacturer which substantially reduce the hydrogen content. A range of mean maximum pore diameters was measured in the wedge with the largest diameters observed in the upper portion of the bottom region. The increase in pore diameter at this location was a result of the double ladle pouring procedure used to fill the casting where the casting began to solidify during the delay between the end of pouring the first ladle and the start of pouring the second.
\subsection{Fatigue test conditions}
All fatigue tests were performed with fully reversed ($R=-1$) loading conditions under load control with a sinusoidal signal. The tension-torsion specimens were loaded in phase on an Instron servo-hydraulic test platform at 11 Hz. The pure torsion and pure tension specimens were tested on a Amsler-Vibraphore machine at 45 Hz. Testing was conducted using the step technique to target the fatigue limit ($\sigma_f,\tau_f$) at 10$^6$ cycles. For this technique, each specimen undergoes cyclic loading at an initial load level estimated based on previous testing experience. Samples that did not fail after 10$^6$ cycles were then cycled again at a higher stress amplitude. The details of the loading conditions and the number of loading steps experience by each sample are listed in Table \ref{table:fatiguespecimens}. For specimens that failed before 10$^6$ cycles without a preceding loading step, the fatigue limit was found using a Basquin coefficient of 0.17 based on experimental results found in the literature \cite{Gao.04,Wang.01,Brochu.10,Jana.10,Ludwig.03}. For specimens that failed before 10$^6$ cycles after a minimum of one loading step, the fatigue limit was calculated as the average stress amplitude of the failure and prior step. 

\begin{table}[h!]

\caption{Test history for all fatigue specimens\label{table:fatiguespecimens}}
    \begin{singlespacing}
        \begin{center}
    \begin{tabular}{ccccccc}
    \hline
    \multicolumn{2}{c}{Specimen}  & \multicolumn{2}{c}{Loading (MPa)} & \multicolumn{2}{c}{Steps} & \multirow{2}{*}{$N_f$ ($\times 10^5$)}\\
    Name & Type${^\dag}$                   &            $\sigma_a$ & $\tau_a$ & Number & MPa/step & \\
                                \hline
    W1 & TT & 0 & 90 & 3 & 5 & 7.22\\
    W2 & TT & 0 & 85 & 1$^{\ddag}$ & N/A & 3.00\\
    W3 & TT & 45 & 78 & 1$^{\ddag}$ & N/A & 8.35\\
    W4 & TT& 90 & 52 & 5 & 5& 1.45\\
    W5 & TT& 70 & 70 & 2 & 5& 1.05\\
    \hline
    B1 & TT& 50 & 87 & 2 & 5& 1.18\\
    B2 & TT& 90 & 52 & 2 & 5& 2.04\\
    B3 & TT& 70 & 70 & 2 & 5& 0.76\\
    B4 & To& 0 & 70 & 2 & 5& 3.98\\
    B5 & TT& 0 & 100 & 3 & 10& 1.51\\
    B6 & TT& 0 & 110 & 2 & 10& 8.83\\
    \hline
    M1 & TT& 50 & 87 & 2 & 5& 2.31\\
    M2 & TT& 90 & 52 & 2 & 5& 4.01\\
    M3 & TT& 95 & 0 & 3 & 5& 0.79\\
    M4 & TT& 65 & 65 & 2 & 5& 2.46\\
    M5 & TT& 70 & 70 & 3 & 5& 5.26\\
    M6 & To& 0 & 60 & 2 & 10 & 3.25\\
    M7 & To& 0 & 55 &1$^{\ddag}$ & N/A & 2.27\\
    M8 & To& 0 & 60 & 4 & 10 & 2.61\\
\hline
    T1 & TT& 65 & 65 & 5 & 5& 4.24\\
    T2 & TT& 65 & 65 & 2 & 5& 1.29\\
    T3 & TT& 65 & 65 & 1$^{\ddag}$ & N/A& 9.08\\
    T4 & TT& 60 & 60 & 1$^{\ddag}$ & N/A&4.05\\
    T5 & Te& 90 & 0 & 5 & 10& 6.63\\
    T6 & To& 0 & 50 &1$^{\ddag}$ & N/A&7.33\\
    T7 & To& 0 & 50 & 2 & 10& 4.84\\
\hline
\end{tabular}
\end{center}
\end{singlespacing}
$^{\dag}$\small{TT: tension-torsion, To: torsion, Te: tension. Tension-torsion specimens W1, W2, B5 and B6 were tested in pure torsion; M3 was tested in pure tension.}\\
$^{\ddag}$\small{Failure before $10^6$ cycles}
\end{table}

\section{Fatigue Test Results \& Criteria Comparison}\label{sec:results_criteria_comp}
The testing conditions and results are summarized in Table \ref{table:fatiguespecimens}. In the majority of the fatigue tests, failure occurred after at least one loading step. However, four specimens failed during the first loading step (Specimens W2, W3, M7, and T6). There were two instances of specimens arriving at the same fatigue limit albeit with different numbers of steps: i) Specimens T1 and T2, both from the top of the wedge, failed at $\sigma_a,\tau_a=65$ MPa after 5 and 2 loading steps, respectively. Specimen T3 employed in the crack growth study (Section \ref{sec:crackpropTensTor}) was conducted as a run-out test and failed after 907 000 cycles at $\sigma_a,\tau_a=65$ MPa. Furthermore, samples W4 and M2 failed at $\sigma_a=90$ and $\tau_a=52$ MPa after 5 and 2 loading steps, respectively. The non-ferrous nature of this material coupled with these limited observations provide speculative insight that the coaxing effect \cite{Nicolas.06,Murakami.84} is not prominent. While the coaxing effect may be present under other loading scenarios, it is asserted that the step testing method was able to ascertain the fatigue limit for $10^6$ cycles to within 5 MPa.

Grouped according to family, Fig 3 depicts $\sigma_f$ versus $\tau_f$ for each specimen. Following the trend of decreasing $\lambda_2$ from top to bottom of the wedge, the general trend is that specimens with the largest $\lambda_2$ exhibit the lowest fatigue limit. The exception was specimens extracted from the wheel, which showed a lower fatigue limit than the wedge material at a tension/torsion ratio slightly below pure torsion. However, this difference is less than the range of results shown by the pure torsion testing of the wheel specimens. Fig 4 depicts $\sigma_a$ versus $\tau_a$ for all of the loading scenarios (Table \ref{table:fatiguespecimens}) independent of family type. Considering the maximum fatigue limit for the entire experimental data set independent of specimen type and microstructure, $\sigma_f$ and $\tau_f$ for $R=-1$ are approximately equidistant from the origin for all ratios of tension to torsion.

\begin{figure}
\centering
\includegraphics[width=0.5\linewidth]{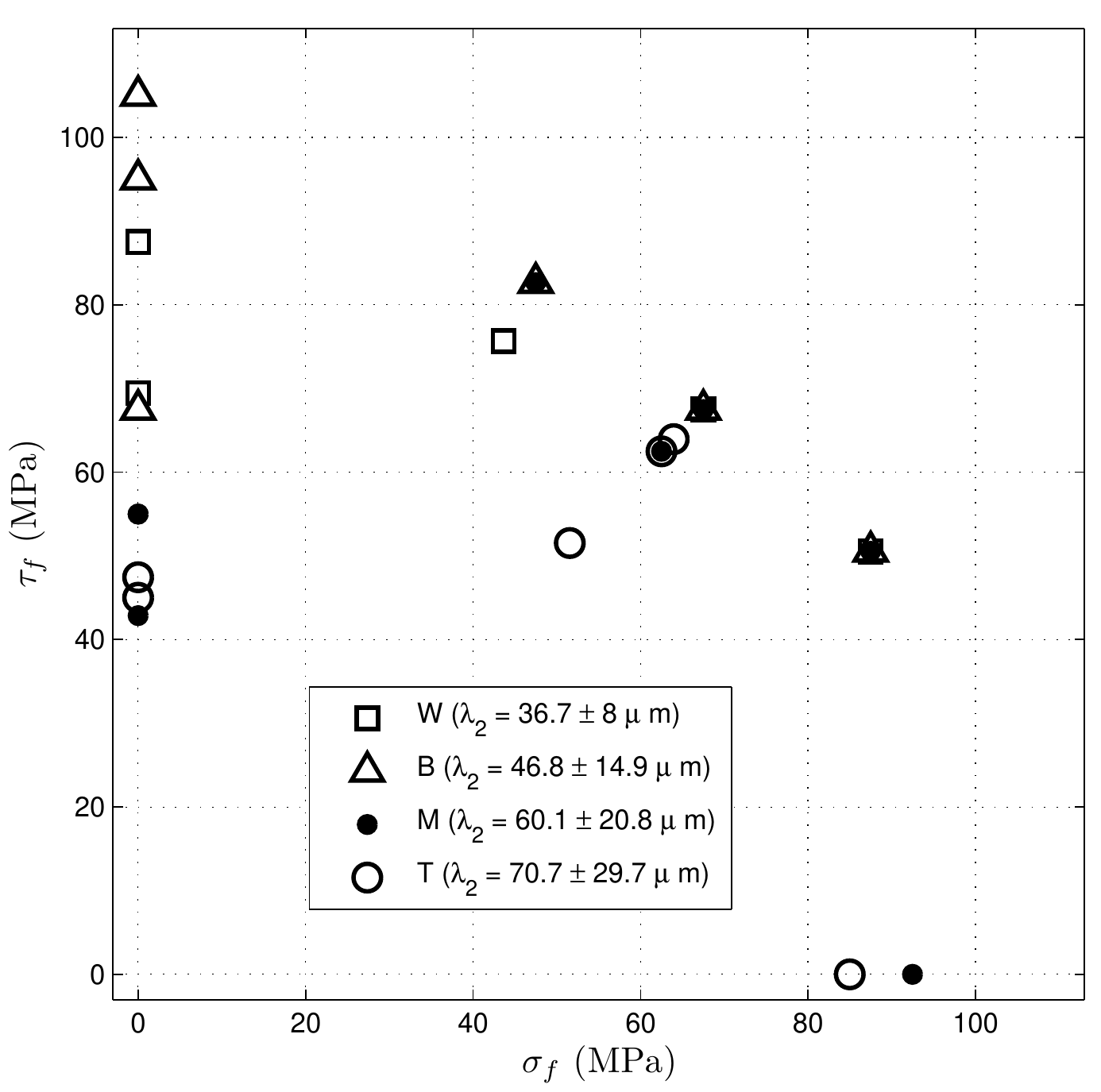}
\caption{Fatigue limit grouped by material family type.}
\end{figure}

Under pure torsion loading, the straight-gauged tension-torsion type specimens demonstrating a higher fatigue limit as compared to torsion-type specimens.  The defect population at different levels of the wedge (Table \ref{table:SDAS_pores}) demonstrate that the largest range of defect sizes was observed in the middle of the casting. This is where $\sim$80 \% of the samples loaded under pure torsion were extracted. During the torsion tests, it was also observed that multiple shear cracks were active at the same time, with the dominant crack not appearing until late in the test. The fractographic observations for this loading condition (Section \ref{sec:loadingMacro}) combined with the porosity measurements preclude specimen configuration being responsible for sample-type variations in fatigue limit under pure torsion.

In order to gauge the existing multiaxial fatigue criteria, the maximum fatigue limit from all data sets (i.e. each specimen family) at each tension/torsion ratio was determined to characterize the behaviour for A356-T6. The extracted fatigue limit data was compared to the Crossland and Maximum Principal Stress (MPS) fatigue criteria. This was done to compare the respective criteria to the entire breadth of experimental results. This is more representative of the microstructural differences in a cast component due to inherent variations in cooling rate. Furthermore, it has been shown that defects play a larger role in determining fatigue resistance as opposed to microstructure \cite{McDowell.03,Roy.11}.

Crossland \cite{Crossland.56} proposed that the second invariant of the deviatoric stress and the maximum hydrostatic stress are the main parameters determining fatigue resilience:
\begin{equation}\label{eq:Crossland}
\sqrt{J_{2,a}}+\rho J_{1_{\max}}\leqslant A
\end{equation}
where $J_{2,a}$ is the amplitude of the second invariant of the deviatoric stress and $J_{1_{\max}}$ is the maximum hydrostatic stress. The constant $\rho$ is a function of the fatigue limit in pure tension and torsion and $A$ is the fatigue limit in pure torsion. Using the average $\sigma_f=89$ MPa and $\tau_f=67$ MPa found in this study, $\rho$ and $A$ were determined to be 0.54 and 67 MPa, respectively.

The MPS criterion asserts that the maximum principal stress must be below a critical threshold such that:
\begin{equation}
\sigma_{1_{\max}}\leq C
\end{equation}
where $C$ is taken to be the average tensile fatigue limit, $\sigma_f=89$ MPa.

These two criteria are plotted versus $\sigma_a$ and $\tau_a$ in Fig 4. Both the Crossland and the MPS criteria underestimate the measured fatigue limit for combined loading.  It should be noted that the Crossland criteria approaches the MPS criteria under tension. As a result, the variance in the pure torsion results may show that the Crossland criteria applied to these results is overly conservative. Since the Crossland criterion is reliant on the determination of an accurate fatigue limit under pure torsion, the MPS criteria is the most conservative criteria to describe these results \cite{Gao.04,Sharma.04}.

\begin{figure}
\centering
\includegraphics[width=0.5\linewidth]{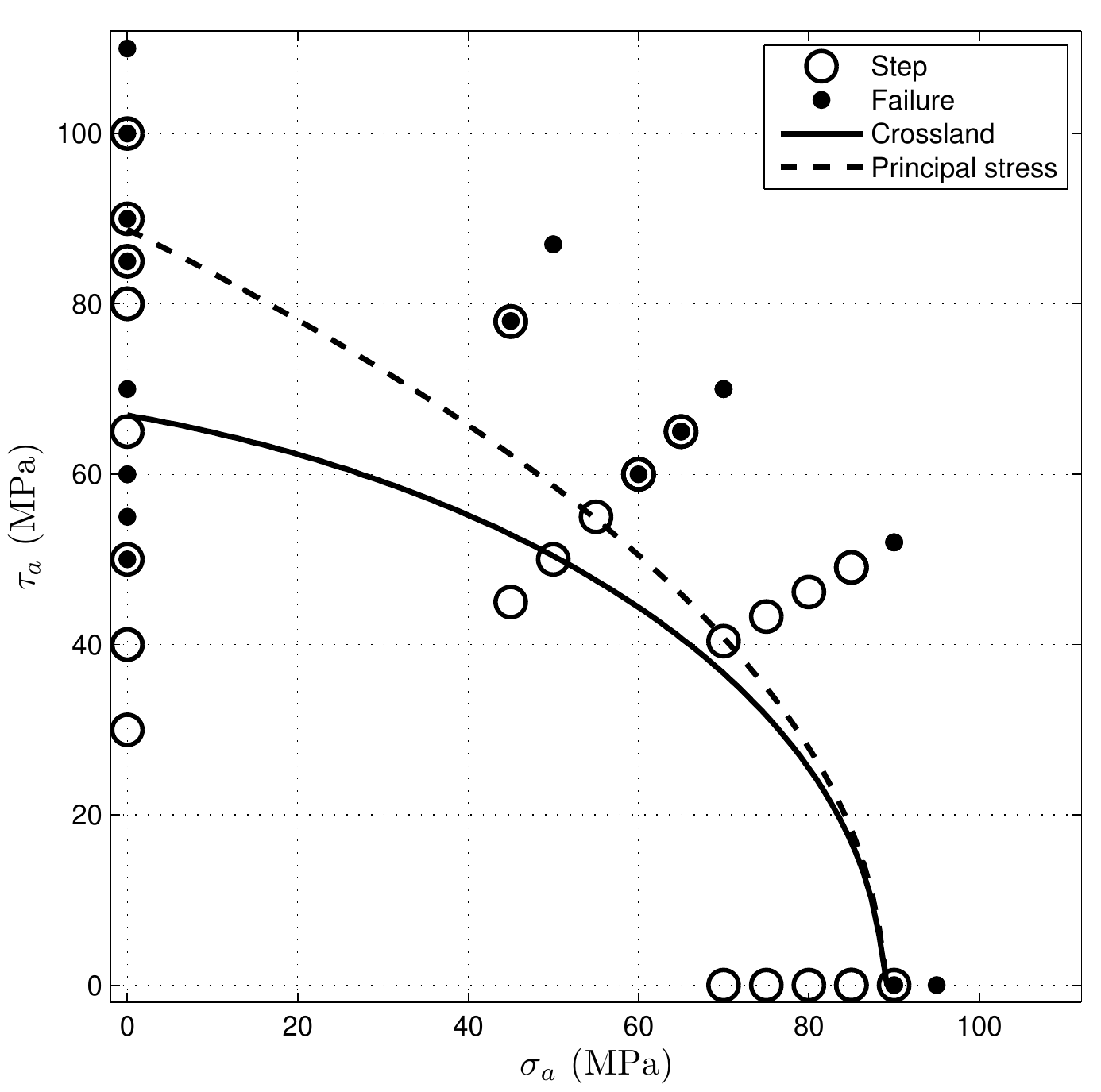}
\caption{Fatigue test points compared to Crossland and Maximum Principal Stress criteria.}
\end{figure}

\section{Fracture surfaces}\label{sec:loadingMacro}
A representative summary of the fracture surfaces formed for each type of loading is presented in Fig 5. Under pure tension (Fig 5a), the fracture plane was found to be always normal to the direction of applied stress and thus, coincident with the maximum principal stress ($\sigma_{1_{\max}}$) plane. Specimens from both the wheel and wedge casting exhibited this behaviour indicating that this observation is independent of microstructure. Furthermore, SEM observations on gage section material far from the initiation site did not reveal any other cracks. Thus, multiple initiation sites did not manifest under pure tension, regardless of microstructure.

\begin{figure}
\centering
\includegraphics[width=0.5\linewidth]{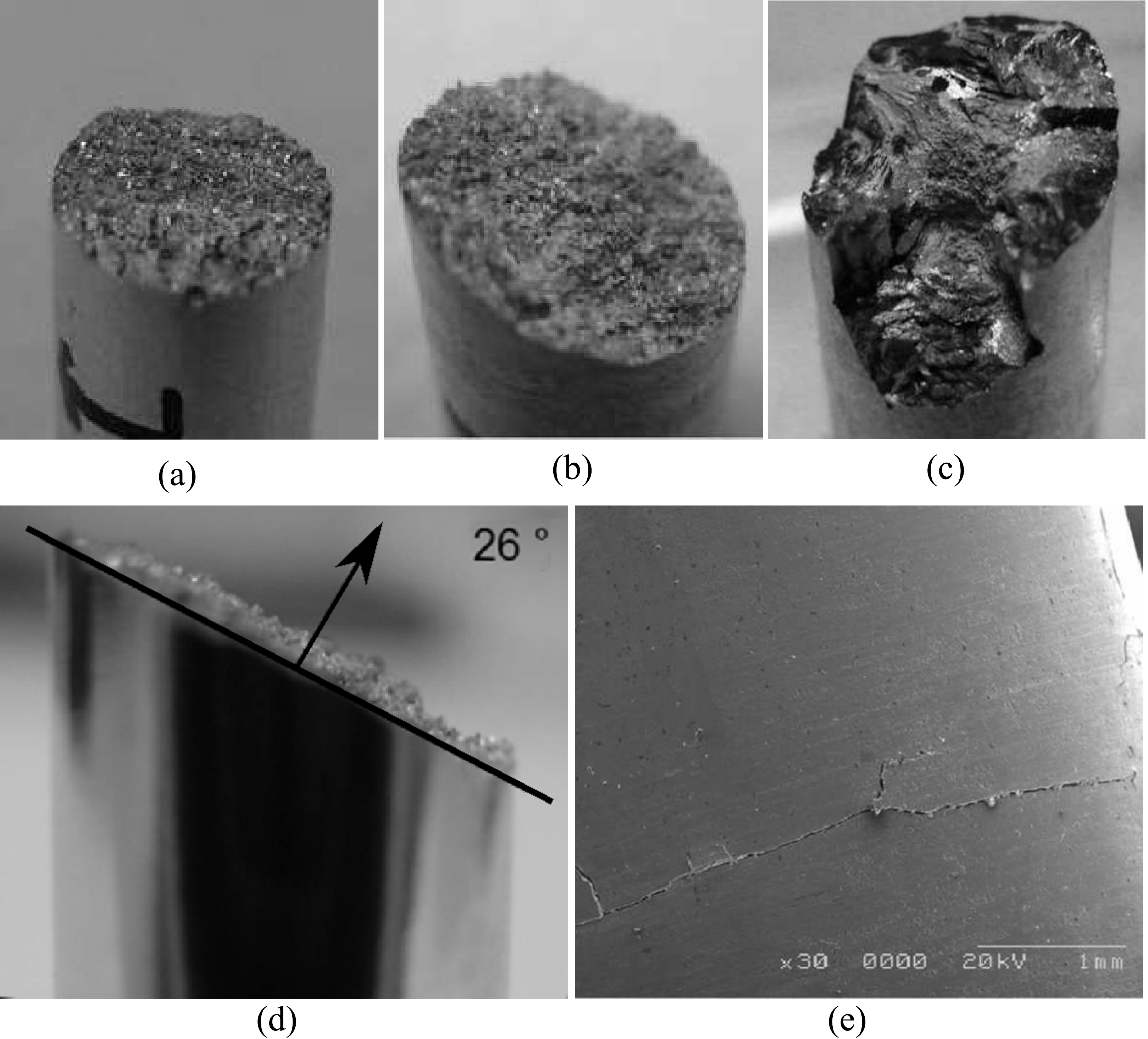}
\caption{Failure types: (a) pure tension (b) combined tension-torsion ($\sigma_a=\tau_a$) (c) pure torsion (d) Macroscopic crack plane orientation combined tension-torsion ($\sigma_a=\tau_a$) (e) shear crack on the gauge section of the torsion sample far from the fracture surface.}
\end{figure}

Under combined tension-torsion loading, the fracture surface features are similar to pure tension including regular fracture planes with no bifurcation (Fig 5b) regardless of the sample location. SEM observation of the gage section far from the fracture surface reveals only small secondary cracks less than 50 $\upmu$m long. The orientation of the fracture surface in Fig 5d is shown by the fracture surface normal. In this combined loading case where $\sigma_a=\tau_a$, the fracture surface normal is oriented 26$^{\circ}$ from the axis of the specimen. With the maximum principal stress acting at 31$^{\circ}$ from the specimen axis, the fracture surface normal is close to being parallel. This difference of $5^{\circ}$ is the largest discrepancy observed over the 18 multiaxial specimens tested. Thus, the orientation of the macroscopic fracture plane under combined loading conditions correlates to the loading condition and more specifically, to the direction of the maximum principal stress ($\sigma_{1_{\max}}$). Therefore, the tension component of the loading must play a major role in deciding the path of crack propagation.

Under pure torsion, the failures surfaces observed were very different and showed no similarities to that of pure tension and combined tension-torsion loading. Fig 5c is an example of the fracture surfaces observed. The tortuous fracture surface has two major crack planes activated: one aligned with the axis of the specimen and the other one perpendicular to this axis. This highlights the difficulty in finding a unique initiation site on the fracture surface. Macroscopic observations confirmed by SEM analysis indicate that there are many different initiation sites over the periphery of the gage section and that different crack planes link together to form the final fracture surface.

Remarkably, there is no evidence of macroscopic cracks growing in the direction normal to the $\sigma_{1_{\max}}$ in the pure torsion scenario. For this loading condition, the crack path is governed by shear as opposed to principal stress. A clear demonstration of the dominance of shear is shown in Fig 5e where cracks propagate in shear mode from early in the fatigue life to the final failure. The very long crack observed in Fig 5e was observed to propagate throughout the test under shear mode III and showed no evidence of bifurcation under mode I.  When  bifurcation did occur on this sample, a new crack plane extended from the original mode III shear plane and linked with another mode III shear crack in the opposing activated shear plane. The sample shown in Fig 5e exhibited more than 10 other cracks similar to the one depicted, and additional smaller cracks observed in both shear mode III planes.

The fractographic observations indicate that the fracture morphology is independent of the sample family. Therefore, the microstructural features and the defect characteristics, such as $\lambda_2$ and average pore diameter (Table \ref{table:SDAS_pores}) do not dictate crack paths under multiaxial loading. The overriding observation from the following analysis is that the macroscopic crack path is governed by $\sigma_{1_{\max}}$ under multiaxial loading except for pure torsion where shear mode III dominates. This observation conflicts with the general mechanical analysis performed in Section \ref{sec:results_criteria_comp} which showed that the MPS criteria provided the closest description of the multiaxial fatigue results even under pure torsion.
\section{Initiation sites}
There are various multi-scale microstructural features that can cause fatigue initiation in A356-T6 \cite{Gao.04,McDowell.03,Yi.03}. Defects such as gas pores, shrinkage pores, oxides and inter-metallic particles can initiate fatigue cracks. At smaller length scales, the fatigue properties are dominated by the $\alpha$-Al and eutectic characteristics. SEM observations were performed on each sample to identify fatigue crack initiation sites and to examine the crack propagation surfaces.

The analysis performed on each sample followed a systematic methodology to reproducibly identify initiation sites and crack surface features.  The methodology employed was as follows:
\begin{itemize}
 \item Use optical microscopy to observe the fracture surface and identify the fatigue and fast-fracture zones. The fast-fracture zone refers to that portion of the crack surface which developed in the last fatigue cycles.
\item Observe the fatigue zone using an SEM to determine the initiation site (1 mm$^2$) where river marks on the fracture surface converge.
\item If a clear defect is identified, measured the size of the defect using the parameter $\sqrt{\text{area}}$ on the fracture surface. This is performed regardless of the position of the defect relative to the gage surface.
    \end{itemize}
In a number of samples, the initiation site could not be accurately identified or characterized. This was true for pure torsion samples with multiple initiation sites, but also for multiaxial tests where model III shear cracking lead to contact and damage of fracture surfaces.

Due to the readily available tensile fatigue data for A356-T6, only two specimens were tested under these conditions. Each of these specimens exhibited a gas or shrinkage pore as the initiation site for the fatal crack. The first, M3 (Table \ref{table:fatiguespecimens}), had a fatigue limit of 90 MPa and contained a gas pore with an equivalent diameter of 88 $\upmu$m at the initiation site. The other specimen, T5, had a shrinkage pore with an equivalent diameter of 370 $\upmu$m (Fig 6a) at the initiation site and exhibited a fatigue limit of 85 MPa.

\begin{figure}
\centering
\includegraphics[width=0.5\linewidth]{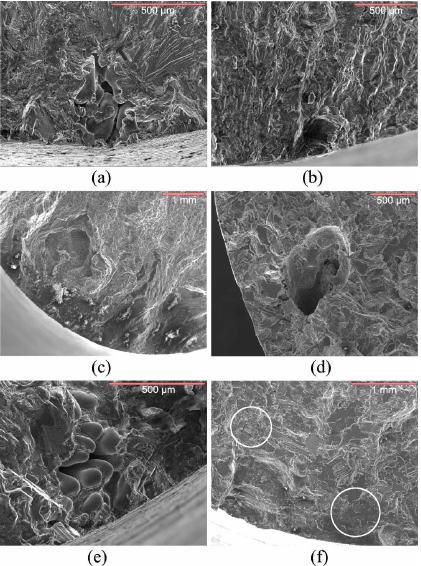}
\caption{Fracture surfaces: (a) specimen T5, $\sigma_f=$ 85 MPa (b) specimen B3, $\sigma_f=\tau_f=$ 68 MPa (c) specimen W5, $\sigma_f=\tau_f=$ 68 MPa (d) specimen M4, $\sigma_f=\tau_f=$ 63 MPa (e) specimen T4, $\sigma_f=\tau_f=$ 51 MPa (f) specimen T7, $\tau_f=$ 45 MPa.}
\end{figure}

Similar to pure tension, the combined tension-torsion samples exhibited fracture surfaces that were easily characterized. Figs 6b-e are examples of typical defects that initiated the fatal crack. Fig 6b shows the fracture surface of sample B3 that had a $\sigma_f$ of 68 MPa. The initiation area on this specimen was readily identified, but a root initiating defect could not be found. Fig 6c is the fracture surface for sample W5, which had the same fatigue limit as B3. The fracture surface is less clear but it was possible to identify the initiation area where no clear defects were observed. The fracture surface instead shows friction-generated oxide associated with crack propagation. Fig 6d reveals a 500 $\upmu$m subsurface pore just below the surface of specimen M4. Remarkably, the fatigue limit for M5 was 63 MPa, which is close to the highest value observed. This may be caused by the root defect location relative to the surface of the sample. When a propagating crack does not intersect the sample surface, it is not under ambient environmental conditions, but under vacuum instead. This effect has been investigated in a cast Al-Si-Cu alloy \cite{Ueno.10} where the fatigue life under vacuum when a surface defect is present is the same as when failure initiates from an internal defect of the same size.  These observations have also been confirmed with nodular cast iron \cite{Billaudeau.04}. Therefore, a fatigue assessment of cast A356  should account for the position of the defect with respect to the free surface as there is different damage accumulation dependant on whether the defect lies on the surface or within the bulk. Fig 6e shows a typical shrinkage pore at the surface of specimen T4. In this case, a 500 $\upmu$m pore decreased the $\sigma_f$ to 51 MPa. This result suggests that the surface shrinkage pore is more detrimental to the fatigue life than the previous subsurface pore of the same size. However, the environmental effect is not the only factor in this case because the morphology of the defects are different.

The fracture surfaces of the samples tested in pure torsion were much more difficult to analyze than the pure tension samples. As shown in Fig 5c, the macroscopic topology of the surface is very complex with two perpendicular mode III cracks activated and multiple initiation sites. Initiation is spatially distributed and as a result, the fracture surface is a combination of different cracks that have coalesced to cause final failure of the sample. Thus, it was difficult to identify a single initiating feature. However, when the fracture surface was less tortuous, there were features that could be analyzed (Fig 6f). For these samples, it was non-trivial to separate the fatigue zone from final failure zone. Fig 6f shows two features that are presumed fatigue initiation sites. The feature closest to the center of the specimen is clearly shrinkage porosity with an equivalent diameter of 265 $\upmu$m, located well away from the surface. The second feature is possibly an oxide film, but the features have been destroyed by damage/oxidation of crack surfaces under mode III propagation.

Under cyclic torsional loading, it is apparent that fatigue mechanisms are related to small, diffuse damage. This suggests that samples tested under pure shear conditions are more susceptible to distributed porosity as compared to the other loading scenarios. The lack of hydrostatic stress in this loading scenario may play a role in shear susceptibility. For steel, it has been shown that pure torsion, or a lack of hydrostatic stress, leads to small distributed shear cracks on the surface of the sample while the other loading states cause more localized fatigue damage \cite{Flaceliere.07,Verreman.07}. However, when a crack in steel initiates on a shear plane, it bifurcates into mode I propagation after a relatively short length (100 - 300 $\upmu$m). In the present study, the pre-bifurcation crack length was much larger (on the order of millimeters as shown in Fig 5e).
\section{Crack propagation}
The propagation of cracks intersecting the surfaces of samples under different loading conditions was studied using the replica technique. In the replica technique, an elastomer is applied to the surface of a specimen at different cycle counts to make a negative imprint of the specimen surface\cite{Palin.02}. The imprints, which show the various stages of crack evolution, are sputter-coated and analyzed via an SEM. In this study, replicas were applied to a tensile specimen that had an artificial defect introduced on its surface. This specimen is independent from those specimens listed in Table \ref{table:fatiguespecimens}. The replica technique was also applied to a combined tension-torsion specimen (T3 in Table \ref{table:fatiguespecimens}) to study natural crack propagation. When translating the replica measurements to the fracture surface, the specimen curvature was taken into account. Crack measurements where the crack length was greater than 75\% of the replica perimeter have been omitted due to replica curvature complications.
\subsection{Crack propagation under pure tension}
The artificial defect specimen was drawn from the bottom of the wedge and had a 416 $\upmu$m defect introduced to the middle of the gage section via Electro-Discharge Machining (EDM). This technique of generating artificial defects has been qualified in other crack propagation investigations \cite{Billaudeau.environmental.04,Nadot.06}. The specimen was then cycled and cycled under pure tension at 90 MPa with the fatal crack localized at the site of the defect. This stress level is comparable to the highest fatigue limit found for the two other samples tested under pure tension (M3 and T5).

The resulting fatigue life of this sample, 673 000 cycles, indicates that the defect had an influence on the fatigue limit. The artificial defect is easily identifiable as the origin of failure on the fracture surface shown in Fig 7. The replica results in Fig 7 show the crack along the surface of the sample at 150 000, 224 000, and 300 000 cycles. Indicators identifying the extent of the crack have been added to the replica images. Through the superposition of the fracture surface and the replicas, it is possible to correlate the crack extent indicators with the fracture surface. The exact configuration of the crack with both the surface and inner shape of the crack front is tractable. As shown by the markers on Fig 7, cracking occurs on either side of the defect and the crack front passes through the defect early in crack propagation.

\begin{figure}
\centering
\includegraphics[width=0.5\linewidth]{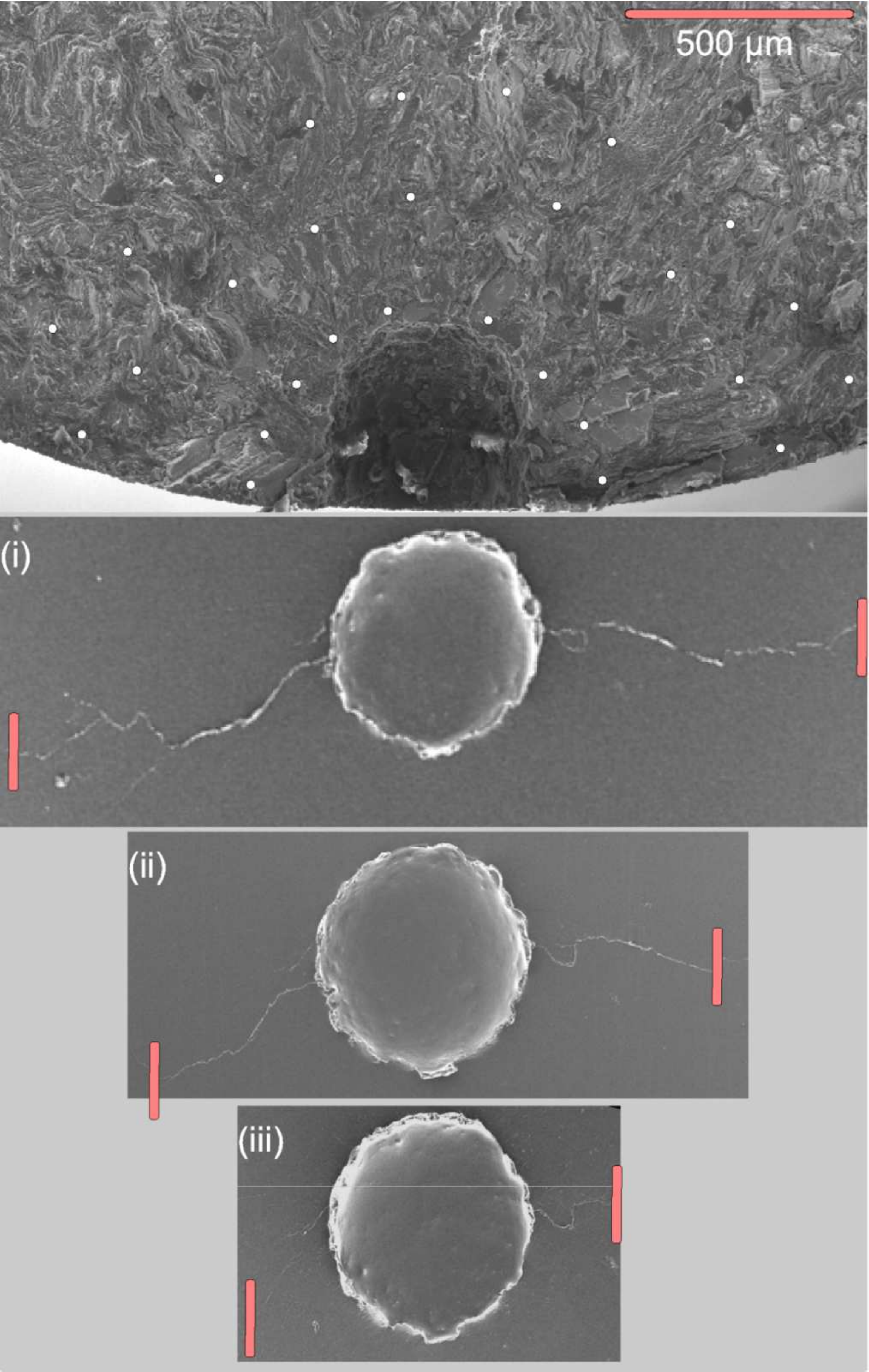}
\caption{Tensile specimen with a 416 $\upmu$m artificial defect, $\sigma_a$= 90 MPa and $N_f = 6.73\times10^5$ cycles. The top image is the fracture surface with crack front marking compared to surface replica at (i) $N = 3\times10^5$, (ii) $N = 2.25\times10^5$ and (iii) $N = 1.50\times10^5$ cycles. Replicas are oriented normal to the loading direction.}
\end{figure}

It is necessary to discard the first stages of crack propagation to study the evolution of an isolated crack far form the influence of the defect. The crack emanating from a spherical defect has a stress intensity factor that is unaffected by its presence at distances greater than 25\% the radius of the defect \cite{Murakami.02,Trantina.84}. Therefore, when the crack is deeper than 520 $\upmu$m, the stress intensity factor can be calculated using Linear Elastic Fracture Mechanics (LEFM). As the crack front geometry is semi-circular, the appropriate shape factor is $2/\pi$. The crack observed on Fig 7 indicates that crack path is perpendicular to the loading direction and locally influenced by the microstructure. Analysis including propagation rates are discussed further in Section \ref{sec:crackprop}.
\subsection{Natural crack propagation under tension-torsion}\label{sec:crackpropTensTor}
Sample T3 was cycled at $\sigma_a = \tau_a=$65 MPa until failure at $N_f=$907 000 cycles. The entire gauge surface of the sample was captured via replicas throughout the test. The replicas revealed that two cracks were present on the gage section at failure. It was also noted that this sample lacked the multiple initiations observed optically on samples following pure torsion loading. The critical crack was located in the middle of the gage section and the secondary crack was at the edge of the gage section. The replica results, showing the evolution of both the primary and secondary cracks, are presented in Fig 8. The crack path is regular and the propagation plane normal of both cracks is oriented at 31$^{\circ}$ from the central axis of the specimen. This is consistent with other tests of the same type (Section \ref{sec:loadingMacro}). Fracture surface analysis revealed a gas pore with an equivalent diameter of 300 $\upmu$m just below the gage surface. Since this specimen had the same fatigue limit (68 MPa) as a specimen where a defect could not be identified, it is assumed that this pore had minimal effect on the fatigue limit. The secondary crack observed on this sample initiated at a gas pore, visible on the surface of the specimen, near the beginning of loading and grew in the same orientation as the main crack.

\begin{figure}
\centering
\includegraphics[width=0.5\linewidth]{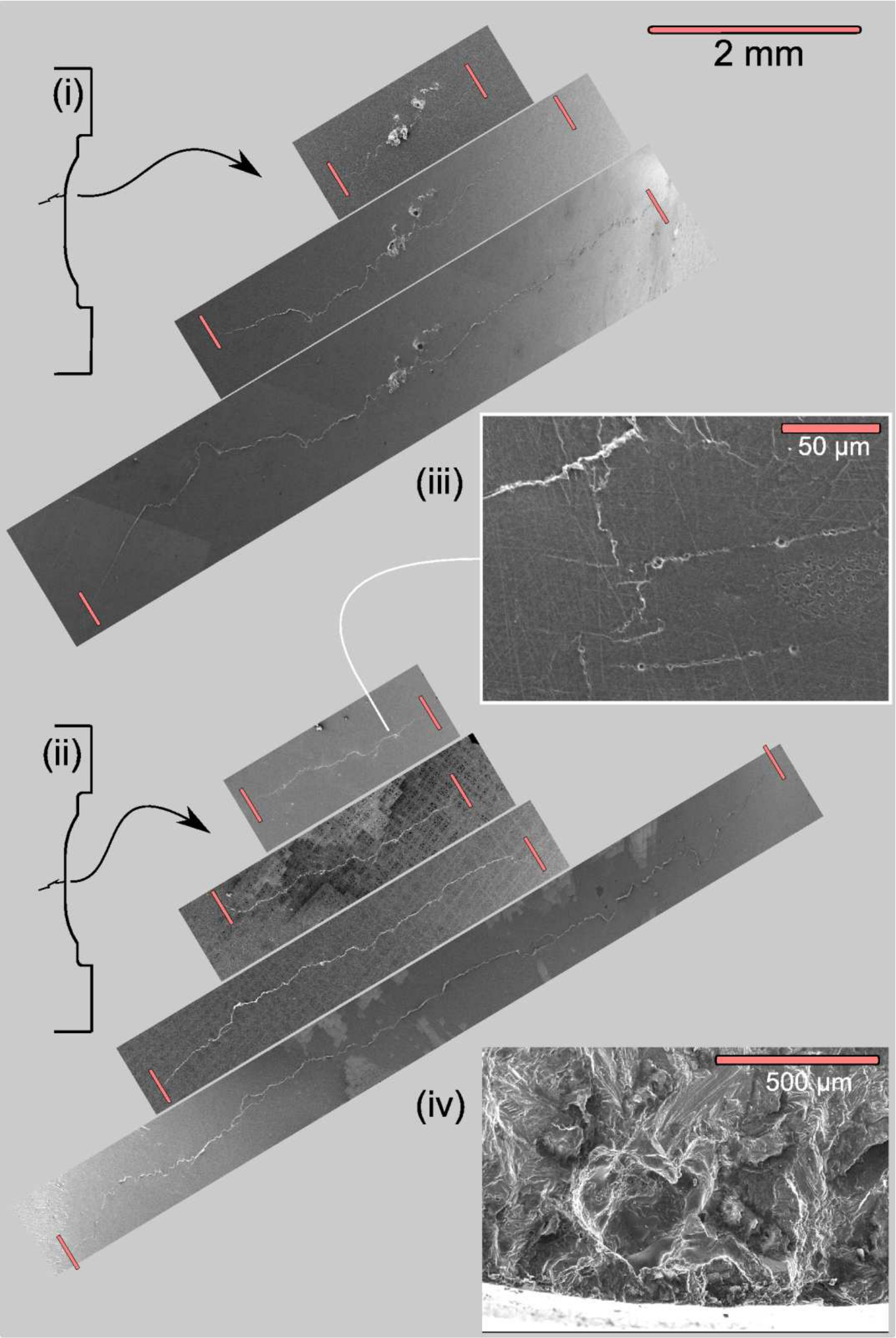}
\caption{Multiaxial specimen T3, $\sigma_a,\tau_a= 65$ MPa and $N_f=9.07\times10^5$ cycles with the secondary crack (i) at $N=2.25\times10^5$, $N=6.75\times10^5$ and $N=8.75\times10^5$. The primary crack (ii) is shown at $N=2.25\times10^5$, $N=3.75\times10^5$, $N=6.75\times10^5$ and $N=8.75\times10^5$. Damage accumulation (iii) was found to occur at the initiation zone and a 300 $\upmu$m defect (iv) was found on the fracture surface.}
\end{figure}

The natural crack growth rate measurements were disregarded until the crack depth was 25\% larger than the radius of the defect, and crack length measurements were only tabulated when the crack depth was greater than 625 $\upmu$m (i.e. the depth of the defect was 500 $\upmu$m from the surface). A similar practice was followed for the secondary crack. Due to the loading type, it was not possible to identify markings on the fracture surface and the exact shape of the crack in the bulk is unknown. However, since the fracture plane is governed by the principal stress, it is assumed that the crack front is semi-circular and identical to the one observed under tension.

It has been observed by others that the first stage of crack propagation links the silicon particles in the brittle eutectic \cite{Gao.04,Buffiere.01,McDowell.03,Ludwig.03}. This is contrary to the findings of other studies \cite{Brochu.10,Gall.00} that have identified the first crack as appearing in the more ductile $\alpha$-Al primary phase. The phase where crack nucleation occurred is not evident as the initiation zone of the replica at high magnification (Fig 8) shows several cracks that are smaller than $\lambda_2$.  However, the very first stages of crack initiation represent a small fraction of the overall fatigue life for HCF and may be therefore discounted since the fatigue life is clearly governed by crack propagation rather than crack initiation.
\subsection{Comparison of natural and long crack propagation}\label{sec:crackprop}
Fig 9 shows the evolution of the crack length $a$ versus the number of loading cycles, where $a$ is half of the crack length measured from the replica. The estimated accuracy of these crack length measurements is  $\pm$5 $\upmu$m. While the crack length data in pure tension is for a test with $\sigma_1=$ 90 MPa, the multiaxial test data is for $\sigma_1=$ 105 MPa. Assuming that $\sigma_1$ is the governing parameter for crack growth, these two tests are not strictly comparable as the stress state for each is different. However, this difference is relatively minor and the natural crack growth behaviour exhibits a very small difference in crack growth versus cycle count despite the different stress states. Therefore, the crack growth data in Fig 9 demonstrates the intrinsic short crack growth behaviour of this alloy. Consequently, the aggregate crack growth data has been used to fit an exponential relationship according to:
\begin{equation}\label{eq:fit}
a=\alpha\exp(\beta N)
\end{equation}
where $2a$ is the surface crack length, $N$ the number of cycles with $\alpha$ and $\beta$ fitting coefficients. As the short crack growth behaviour shows inherent variations, this exponential fit characterizes the average short crack growth curve with $\alpha=470.4$ and $\beta=2.32\times 10^{-6}$ (Fig 9). The crack evolution described by Eq. \ref{eq:fit} lacks the non-linear behaviour associated with initiation. However, based on the degree this equation represents the crack growth data, it is apparent that propagation dominates the fatigue life for A356-T6. The investigation of precise correlations for the initiation stage of the fatigue life of A356-T6 has demonstrated that fatigue initiation can be neglected for pure tension\cite{McDowell.03}. For fatigue lives close to 1 million cycles, the present findings assert that fatigue initiation may be considered negligible regardless of loading condition.

\begin{figure}
\centering
\includegraphics[width=0.5\linewidth]{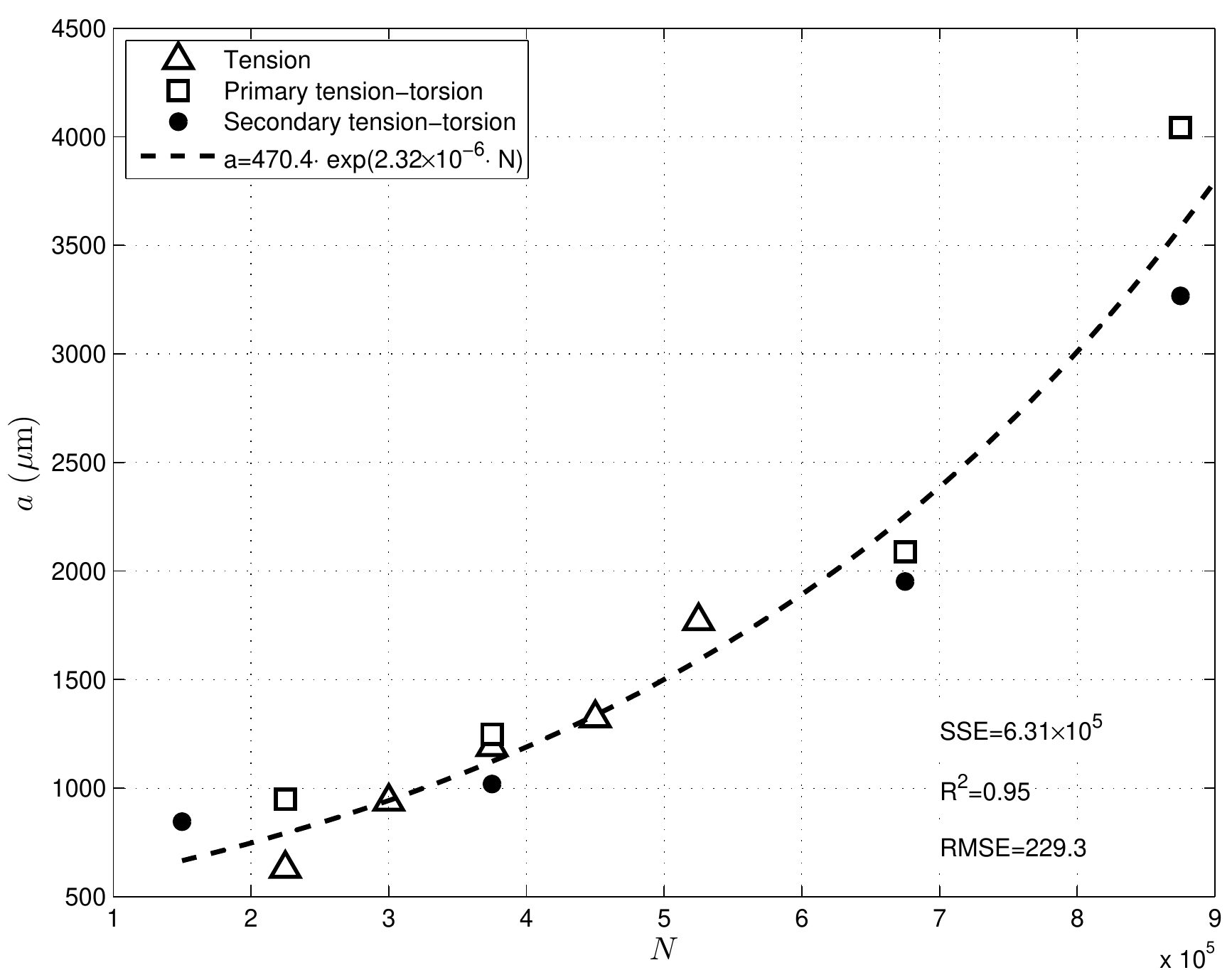}
\caption{Crack growth law under tension and tension-torsion.}
\end{figure}

The crack growth rates are plotted against the amplitude of the stress intensity factor in Fig. 10. The crack growth rates for each stress state were determined from the fitted data (Eq. \ref{eq:fit}). The stress intensity factor was computed assuming a semi-circular crack front, where $K=\phi\sigma_1\sqrt{\pi a}$ with $\phi=0.643$ \cite{Billaudeau.environmental.04,Nadot.99}. The results for the two stress states are compared against studies performed on A319 \cite{Dabayeh.96} by Dabayeh et al., B319-T6 and unmodified A356-T6\cite{Chan.03} by Chan et al., and peak-aged A356-T6 \cite{Skallerud.93} by Skallerud et al. \cite{Skallerud.93}. These four data sets for similar materials with different load ratios are plotted to provide references for short, long, and unstable crack growth behaviours.

\begin{figure}
\centering
\includegraphics[width=0.5\linewidth]{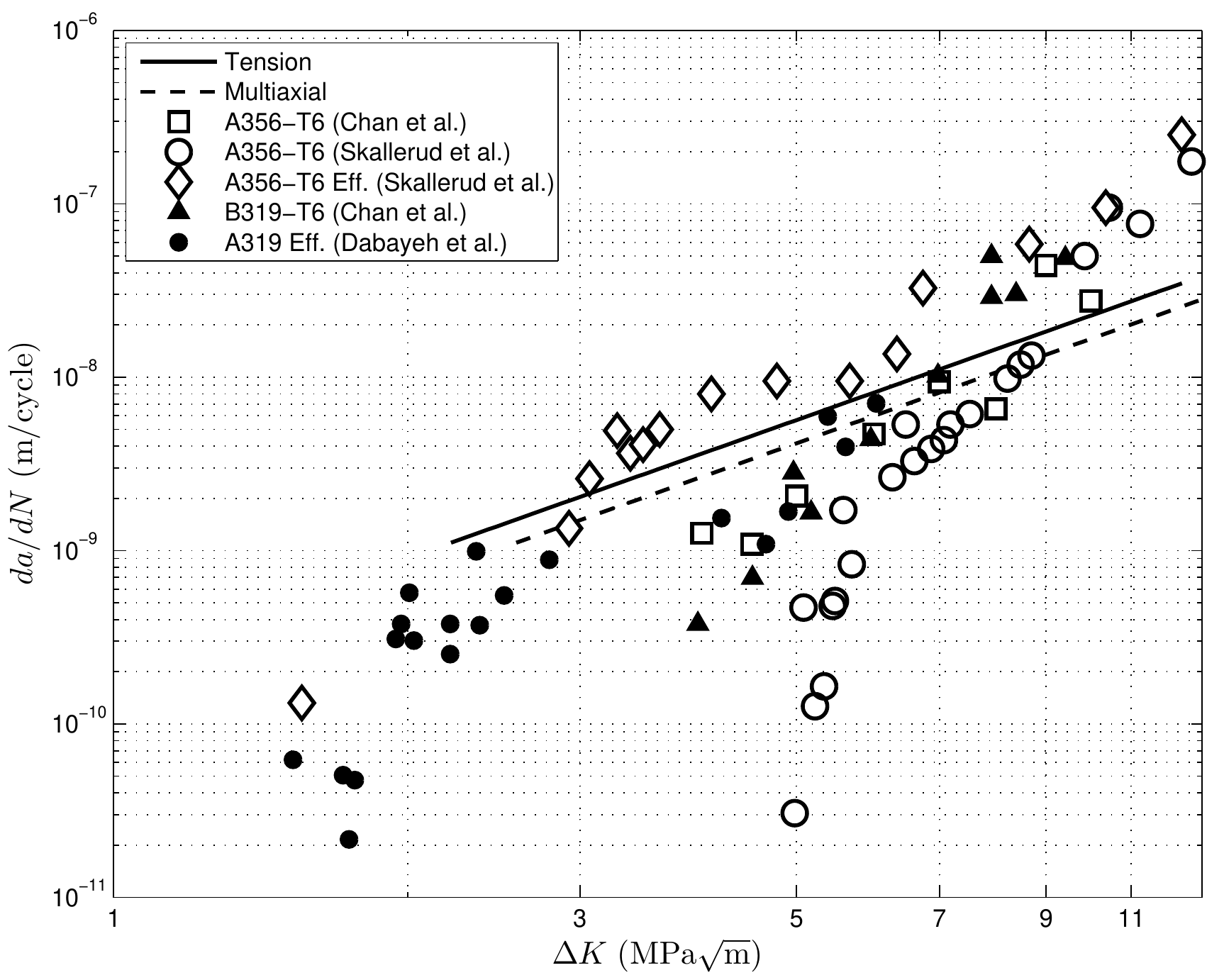}
\caption{Comparison between effective long crack growth behaviour and natural crack growth behaviour.}
\end{figure}

Dabayeh et al. and Skallerud et al. found the crack opening stress intensity $K_{\mathrm{op}}$ needed to attain similar $\Delta K_{\mathrm{eff}}$ for A319 and A356 respectively. The data from Skallerud et al. clearly shows the difference between $\Delta_K$ and $\Delta K_{\mathrm{eff}}$ for A356-T6. The tensile and multiaxial crack growth rates from the current study are approximately the same as that of $\Delta K_{\mathrm{eff}}$ Skallerud et al., which reinforces the short crack categorization microstructure, as it has been found appropriate to employ a closure-free $K$ value to describe short crack behaviour \cite{Miller1.87,Miller2.87}.

The long crack growth data for A319 shows the same effective long crack threshold for A356-T6 as the reported effective long crack threshold is equal to 1.5 MPa $\sqrt{\text{m}}$ in A356-T6 \cite{Zhu.07,Kumar.10,Gall.00}. This indicates that the natural crack growth rate is similar to the effective long crack growth rate. This is contrary to the short crack effect observed in many metallic materials \cite{Suresh.84,Lindley.95}, where effective long crack growth requires a completely positive load cycle ($R > 0$). The natural cracks in this study were cycled with fully reversed loading ($R = -1$) conditions.

With the exception of the pure torsion results, all of the preceding results and observations have indicated that the principal stress is the governing mechanical parameter for crack propagation even under multiaxial loading. Therefore, the Paris law may be used to describe natural crack growth in A356-T6 under multi-axial loading as:
\begin{equation}
\frac{da}{dN}=2 \times 10^{-10}\Delta K^2
\end{equation}
where ${da}/{dN}$ is the crack growth rate in meters per cycle and $\Delta K$ is the positive component of the load cycle in MPa $\sqrt{\text{m}}$. As ${da}/{dN}$ versus $\Delta K$ for both natural cracks analyzed via the replica technique (Fig 10) show good agreement with aggregate long crack data for similar alloys, the MPS assertion is further validated.
\section{Summary and Conclusions}
Employing the step technique, A356-T6 specimens with a wide range of microstructure were submitted to multiaxial HCF loading. The application of various fully reversed tension-torsion ratios generated fatigue limits that were compared to classical fatigue criteria. Fatigue cracks were found to initiate either on casting defects or inside the microstructure. Both scales are in competition for the localization of cyclic plastic deformation that induces the initiation of the crack that leads to failure. When the crack initiates on a defect, it is typically of two different types: oxides or pores (gas or shrinkage). The distance to the free surface as well as the morphology are important parameters. This study has provided the following conclusions:
\begin{itemize}
\item Standard criteria like Maximum Principal Stress (MPS) and Crossland provide mostly conservative estimates of the experimental fatigue limits. However, MPS provides the closest fit to the mechanical results.
\item Cracking mechanisms are very different depending on the loading type. Under pure torsion the material shows multiple initiation sites and long mode III/I shear crack propagation. The bifurcation to mode I is not observed at the end of the fatigue life. Under tension or combined loading the initiation is shorter and more localized such that crack propagation starts directly in mode I.
\item For one million cycles, the fatigue life is governed by crack propagation rather than initiation life regardless of the loading condition.
\item The natural crack growth rate has been found to be the same as for long cracks for similar materials.
\end{itemize}
\singlespacing
\newcounter{Ref}
\setcounter{Ref}{0}

\end{document}